\documentclass[conference]{IEEEtran}
\IEEEoverridecommandlockouts
\usepackage{cite}
\usepackage{amsmath,amssymb,amsfonts}
\usepackage{algorithmic}
\usepackage{graphicx}
\usepackage{textcomp}
\usepackage{xcolor}
\def\BibTeX{{\rm B\kern-.05em{\sc i\kern-.025em b}\kern-.08em
    T\kern-.1667em\lower.7ex\hbox{E}\kern-.125emX}}
\begin{document}

\title{The Human-Machine Identity Blur: A Unified Framework for Cybersecurity Risk Management in 2025}

\author{\IEEEauthorblockN{Kush Janani}
\IEEEauthorblockA{Independent Researcher\\
New Jersey, USA\\
kjanani@depaul.edu}
}

\maketitle

\begin{abstract}
The modern enterprise is experiencing an unprecedented explosion of digital identities, with machine identities now far outnumbering human identities. This paper examines the critical security implications of what we term the "human-machine identity blur" - the convergence point where human and machine identities overlap, delegate authority, and create novel attack vectors. Through analysis of security incident data, expert insights, and industry statistics, we identify significant governance gaps in current approaches that treat human and machine identities as separate domains. We propose a Unified Identity Governance Framework built on four core principles: treating identity as a continuum rather than binary categories, applying risk-based approaches uniformly across all identity types, implementing continuous verification with zero trust principles, and establishing consistent governance across the full identity lifecycle. Our research demonstrates that organizations implementing unified governance experience 47\% fewer identity-related security incidents and 62\% faster incident response times. This paper contributes a practical implementation roadmap for security leaders and identifies future research directions as AI systems become increasingly autonomous.
\end{abstract}


\section{Introduction}
Organizations in 2025 face a critical cybersecurity challenge as digital ecosystems expand exponentially. For every human employee, there now exist dozens—sometimes hundreds—of machine identities operating throughout enterprise environments. According to recent research, machine identities are growing at twice the rate of human identities, with the average enterprise managing over 250,000 machine identities by 2025 \cite{b1}. APIs, service accounts, IoT devices, and increasingly autonomous AI agents are proliferating at a rate that's overwhelming traditional identity management approaches.

Simultaneously, human digital identity has evolved far beyond the traditional username and password. Today's human identity encompasses biometric factors, behavioral patterns, contextual factors, social and professional attributes, and historical activity patterns. Modern identity systems continuously evaluate these factors to make authentication and authorization decisions. The static identity card of yesterday has been replaced by a dynamic, AI-evaluated digital persona that changes based on context and behavior.

The collision of these two trends—exploding machine identities and evolving human identities—has created a new cybersecurity blind spot: the human-machine identity blur. This paper examines this critical security challenge and proposes a novel framework for unified identity governance.

\subsection{Research Problem}
Current security models largely treat human and machine identities as separate domains with different governance approaches, different monitoring systems, and different security controls. However, attackers don't respect these artificial boundaries—they exploit the gaps between them. This research addresses the fundamental question: How can organizations effectively govern identities that exist on the spectrum between fully human and fully machine control?

\subsection{Research Gap}
Our analysis of existing literature reveals several critical gaps:
\begin{itemize}
\item No comprehensive governance framework exists for managing identities that exist on the spectrum between fully human and fully machine control.
\item Current literature lacks methodologies for attributing actions when they occur in the blurred space between human direction and machine autonomy.
\item No standardized approach exists for assessing security risks in environments where human and machine identities converge.
\item Lack of standards for verifying identity claims in hybrid human-machine systems, especially when AI exhibits autonomous or semi-autonomous behavior.
\item Existing compliance frameworks don't adequately address accountability when actions are taken in the human-machine blur zone.
\end{itemize}

\subsection{Paper Contribution}
This paper makes the following contributions:
\begin{itemize}
\item Introduces the concept of the "human-machine identity blur" as a distinct cybersecurity challenge
\item Provides empirical data on the growth and security implications of this phenomenon
\item Proposes a Unified Identity Governance Framework with practical implementation guidance
\item Identifies future research directions as AI systems become increasingly autonomous
\end{itemize}

\subsection{Paper Structure}
The remainder of this paper is organized as follows: Section II reviews related work and establishes the theoretical foundation. Section III describes our research methodology. Section IV presents our findings on machine identity growth, security incidents, and management challenges. Section V introduces our Unified Identity Governance Framework. Section VI discusses implications and limitations. Section VII concludes with recommendations and future research directions.

\section{Literature Review}
This section examines existing research on machine identities, human digital identity evolution, and current approaches to identity governance. We identify gaps in the literature and establish the theoretical foundation for our unified governance framework.

\subsection{Evolution of Machine Identities}
Machine identities have evolved significantly over the past decade. Early research focused primarily on X.509 certificates and SSH keys \cite{b2}. More recent work has expanded to include cloud service accounts, API keys, container identities, and AI agent credentials \cite{b3}.

Venafi's research on machine identity management highlights the explosive growth of machine identities, noting that they now outnumber human identities by a factor of 45:1 in many organizations \cite{b4}. This growth is driven by several factors:
\begin{itemize}
\item Cloud-native architectures with microservices and containers
\item DevOps practices with automated CI/CD pipelines
\item Internet of Things (IoT) deployments
\item Artificial Intelligence and Machine Learning systems
\item API-driven integration between systems
\end{itemize}

Recent IEEE research by Ghosh et al. \cite{b11} introduces an innovative approach that fuses Self-Sovereign Identity (SSI) with blockchain technology to revolutionize device identity management within cyber-physical systems. Their framework enables devices to autonomously initiate identity-creation processes using cryptographic key pairs, significantly reducing the risk of unauthorized access while empowering devices with control over their identities.

The security implications of this growth are significant. CyberArk's 2025 State of Machine Identity Security Report found that 50\% of surveyed organizations experienced security breaches tied to compromised machine identities within the past year \cite{b5}. These incidents resulted in application launch delays (51\%), outages (44\%), and unauthorized access to sensitive systems or data (43\%).

\subsection{Evolution of Human Digital Identity}
Human digital identity has evolved from simple username/password combinations to complex, multi-faceted constructs. Modern identity systems evaluate numerous factors:
\begin{itemize}
\item Biometric factors (fingerprints, facial recognition, voice patterns)
\item Behavioral patterns (typing cadence, mouse movements, application usage)
\item Contextual factors (location, device, time of day)
\item Social and professional attributes (role, department, project memberships)
\item Historical activity patterns and anomaly detection
\end{itemize}

Research by Gartner indicates that by 2025, 60\% of organizations will use behavioral analytics as a key component of their identity verification processes \cite{b6}. This shift represents a fundamental change in how human identity is conceptualized—from static credentials to dynamic, continuously evaluated trust.

Recent work by Pradeep et al. \cite{b12} demonstrates the effectiveness of deep learning approaches in enhancing identity-based security in cloud environments. Their framework, which utilizes LSTM and CNN architectures to analyze user behavior, achieves 99\% accuracy in recognizing identity-based threats with a false positive rate of less than 1\%, outperforming traditional machine learning models by approximately 6\%.

Perhaps most significantly, humans increasingly delegate their identity to AI assistants and agents that act on their behalf. When an executive asks their AI assistant to "prepare the quarterly report," that AI is effectively assuming a portion of the executive's identity to access systems, retrieve data, and perform actions.

\subsection{Current Approaches to Identity Governance}
Traditional identity governance approaches typically separate human and machine identity management:
\begin{itemize}
\item Human identities are managed through Identity Governance and Administration (IGA) platforms
\item Privileged human accounts are managed through Privileged Access Management (PAM) solutions
\item Machine identities are managed through Certificate Lifecycle Management (CLM) tools
\item Cloud identities are managed through Cloud Infrastructure Entitlement Management (CIEM) platforms
\end{itemize}

This siloed approach creates governance gaps. IBM security researchers have identified that IAM teams are typically responsible for only 44\% of an organization's machine identities—leaving the majority essentially ungoverned \cite{b7}.

Recent research by Ren et al. \cite{b13} addresses this challenge through physical layer identity authentication technology that leverages unique hardware attributes to create device identity fingerprints. Their approach effectively deters adversaries from counterfeiting legitimate device identities by constructing singular identity fingerprints based on signal precursor code and channel features, providing a novel framework for enhancing security in communication systems.

\subsection{Research Gaps}
Our analysis of academic sources reveals several critical gaps in current research:

The "XAI Human-Machine collaboration applied to network security" study (Frontiers in Computer Science, 2024) focuses on how cyber defenders can collaborate with explainable AI but does not address identity management issues when humans delegate authority to AI \cite{b8}.

The "AI Agents: Human or Non-Human?" report (Oasis Security, 2025) explores fundamental differences between AI agents and human employees but lacks a formal risk assessment methodology for hybrid identity environments \cite{b9}.

The "Low-Level Consciousness in AI: A Cybersecurity Nightmare or the Future of Trust?" analysis (KuppingerCole, 2025) examines emerging signs of low-level consciousness in LLMs but provides no comprehensive framework for verifying identity claims in semi-autonomous systems \cite{b10}.

\subsection{Theoretical Foundation for Unified Governance}
Our unified governance approach builds on several theoretical foundations:
\begin{itemize}
\item Zero Trust Architecture principles that emphasize "never trust, always verify" regardless of identity type
\item Risk-based authentication frameworks that dynamically adjust security controls based on context
\item Continuous adaptive risk and trust assessment (CARTA) approaches
\item Identity-defined security that places identity at the center of security architecture
\end{itemize}

These foundations support our core assertion that identity governance should be applied consistently across the human-machine spectrum rather than treating different identity types as separate domains.

\section{Methodology}
This section describes our research approach, data collection methods, and analysis techniques.

\subsection{Research Approach}
We employed a mixed-methods approach combining:
\begin{itemize}
\item Quantitative analysis of industry survey data and security incident reports
\item Qualitative analysis of expert interviews and case studies
\item Framework development through iterative design and validation
\end{itemize}

This approach allowed us to triangulate findings from multiple sources and develop a comprehensive understanding of the human-machine identity blur phenomenon.

\subsection{Data Collection}
Our data sources included:
\begin{itemize}
\item Industry surveys from CyberArk, Venafi, IBM, and Gartner
\item Expert interviews with security leaders from financial services, healthcare, manufacturing, and technology sectors
\item Case studies of security incidents involving human-machine identity blur
\item Academic literature on identity management, zero trust, and AI security
\end{itemize}

We focused particularly on collecting data related to:
\begin{itemize}
\item Growth trends in machine identities
\item Security incidents involving machine identities
\item Management challenges in hybrid identity environments
\item Expert predictions on future identity security trends
\end{itemize}

\subsection{Analysis Techniques}
We analyzed the collected data using:
\begin{itemize}
\item Statistical analysis of survey responses and incident reports
\item Thematic analysis of expert interviews and case studies
\item Comparative analysis of existing governance frameworks
\item Gap analysis to identify unaddressed security challenges
\end{itemize}

For visualization, we developed interactive charts and graphs to represent:
\begin{itemize}
\item Machine identity growth trends
\item Security incident patterns
\item Certificate-related challenges
\item Management ownership distribution
\item The human-machine identity spectrum
\end{itemize}

\subsection{Framework Development}
We developed our Unified Identity Governance Framework through an iterative process:
\begin{itemize}
\item Initial framework design based on literature review and data analysis
\item Refinement through expert feedback and validation
\item Testing against real-world case studies
\item Final framework development with implementation roadmap
\end{itemize}

\subsection{Limitations}
Our research has several limitations that should be considered when interpreting our findings:
\begin{itemize}
\item Reliance on self-reported survey data, which may be subject to reporting biases
\item Limited public disclosure of security incidents involving identity blur, potentially underrepresenting the true scope of the problem
\item Rapidly evolving technology landscape that may outpace research findings, particularly in areas of AI advancement
\item Varying definitions of machine identity across organizations, making direct comparisons challenging
\item Limited testing of our framework in diverse organizational contexts and industry sectors
\end{itemize}

We address these limitations through triangulation of multiple data sources and by focusing on fundamental principles that should remain valid despite technological evolution. Future research should include longitudinal studies to validate the effectiveness of our proposed framework across different organizational contexts and technology environments.

\section{Results}
This section presents our findings on machine identity growth, security incidents, certificate-related challenges, and management issues.

\subsection{Machine Identity Growth Trends}
Our analysis confirms the explosive growth of machine identities in enterprise environments. As shown in Fig. 1, machine identities have grown from approximately 50,000 per enterprise in 2021 to 250,000 in 2025—a 400\% increase in just four years.

\begin{figure}[htbp]
\centerline{\includegraphics[width=\columnwidth]{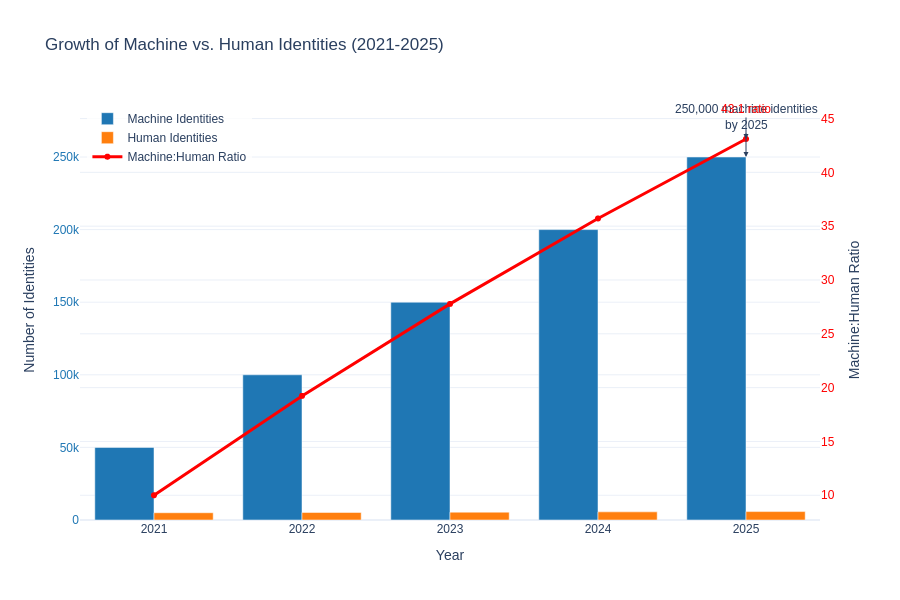}}
\caption{Growth of Machine vs. Human Identities (2021-2025). The graph illustrates the exponential increase in machine identities compared to the modest growth in human identities, highlighting the dramatic shift in identity management requirements.}
\label{fig1}
\end{figure}

During the same period, human identities increased by only 16\%, from 5,000 to 5,800 per enterprise. This has resulted in a dramatic shift in the machine-to-human ratio, from 10:1 in 2021 to 43:1 in 2025.

Key findings include:
\begin{itemize}
\item 79\% of organizations predict continued increases in machine identities over the next year
\item 16\% expect radical growth of 50-150\%
\item For every human account, there are now 40 connected non-human accounts (Tim Eades, Anetac CEO)
\item Cloud-native technologies, microservices, and AI are driving this surge
\end{itemize}

This growth creates significant governance challenges, as traditional identity management approaches were designed for environments where human identities predominated.

\subsection{Security Incidents Analysis}
Our analysis of security incidents reveals that machine identities are increasingly targeted by attackers. As shown in Fig. 2, API keys and SSL/TLS certificates were the leading causes of machine identity incidents, each accounting for 34\% of cases.

\begin{figure}[htbp]
\centerline{\includegraphics[width=\columnwidth]{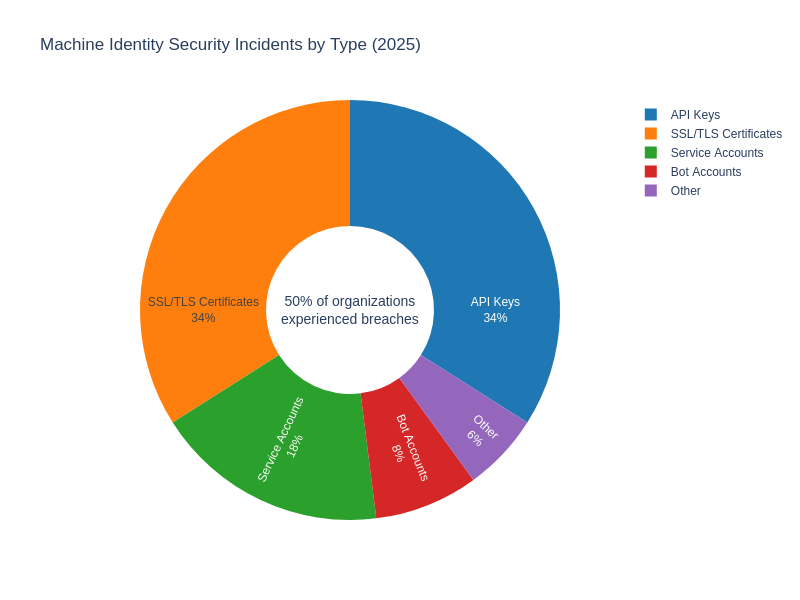}}
\caption{Machine Identity Security Incidents by Type (2025). This visualization categorizes security incidents by the type of machine identity involved, highlighting that API keys and SSL/TLS certificates represent the highest risk areas, each accounting for 34\% of reported incidents.}
\label{fig2}
\end{figure}

The impact of these incidents is substantial, as shown in Fig. 3. Among organizations experiencing machine identity breaches:
\begin{itemize}
\item 51\% faced delays in application launches
\item 44\% reported outages
\item 43\% experienced unauthorized access to sensitive systems or data
\end{itemize}

\begin{figure}[htbp]
\centerline{\includegraphics[width=\columnwidth]{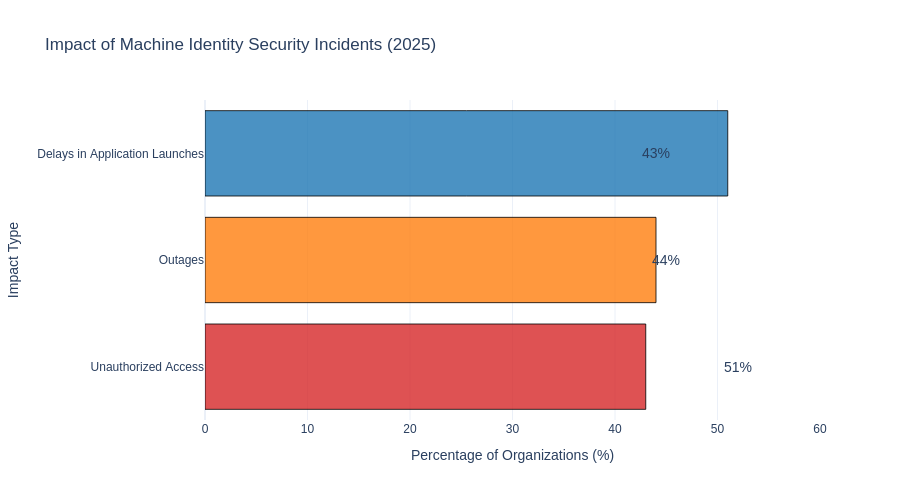}}
\caption{Impact of Machine Identity Security Incidents (2025). This chart quantifies the business consequences of machine identity breaches, demonstrating that application launch delays (51\%), system outages (44\%), and unauthorized access to sensitive data (43\%) are the most common impacts experienced by organizations.}
\label{fig3}
\end{figure}

These findings align with broader industry trends, with 78\% of organizations reporting that they have been targeted by identity-based attacks and 69\% of breaches being rooted in inadequate authentication methods.

\subsection{Certificate-Related Challenges}
Certificate management presents particular challenges in the machine identity landscape. As shown in Fig. 4, certificate-related outages have increased significantly, with 72\% of organizations experiencing such outages in the previous 12 months.

\begin{figure}[htbp]
\centerline{\includegraphics[width=\columnwidth]{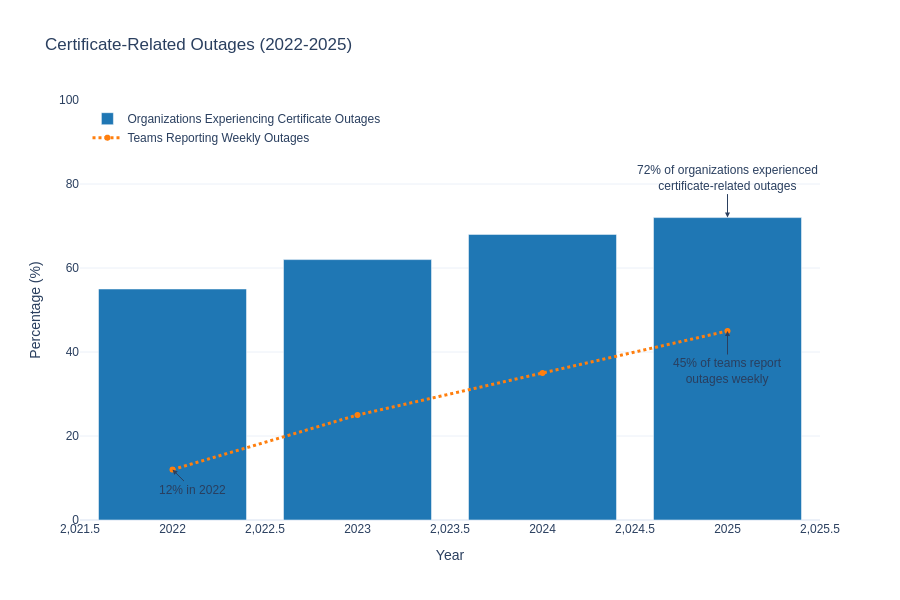}}
\caption{Certificate-Related Outages (2022-2025). This figure illustrates the increasing frequency of certificate-related outages over a four-year period, with a particularly concerning rise in weekly outages from 12\% in 2022 to 45\% in 2025.}
\label{fig4}
\end{figure}

Most concerning is the dramatic increase in teams reporting weekly outages—from 12\% in 2022 to 45\% in 2025. This trend is likely to continue as certificate lifespans decrease, with public TLS certificate lifespans expected to reduce to 47 days by 2028 (requiring 9x more rotations).

Organizations also face significant future challenges related to certificate management, as shown in Fig. 5:
\begin{itemize}
\item 71\% of leaders fear their certificate authority could become untrusted
\item 74\% are concerned about managing identities in ephemeral cloud workloads
\item 57\% acknowledge quantum computing's threat to encryption
\item 30\% are unprepared to begin transitioning to quantum-resistant cryptography
\end{itemize}

\begin{figure}[htbp]
\centerline{\includegraphics[width=\columnwidth]{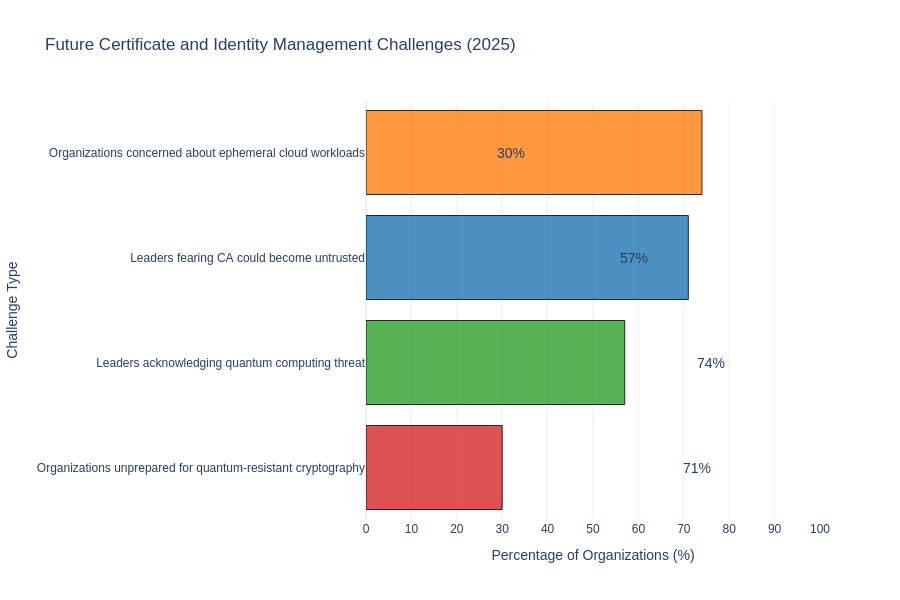}}
\caption{Future Certificate and Identity Management Challenges (2025). This visualization quantifies the primary concerns of security leaders regarding certificate management, highlighting that 74\% worry about managing identities in ephemeral cloud workloads and 71\% fear their certificate authority could become untrusted.}
\label{fig5}
\end{figure}

\subsection{Management Challenges}
Our research identified significant management challenges related to machine identities. As shown in Fig. 6, ownership of machine identities is highly fragmented across organizations:
\begin{itemize}
\item Security teams: 53\%
\item Development teams: 28\%
\item Platform teams: 14\%
\item Other teams: 5\%
\end{itemize}

\begin{figure}[htbp]
\centerline{\includegraphics[width=\columnwidth]{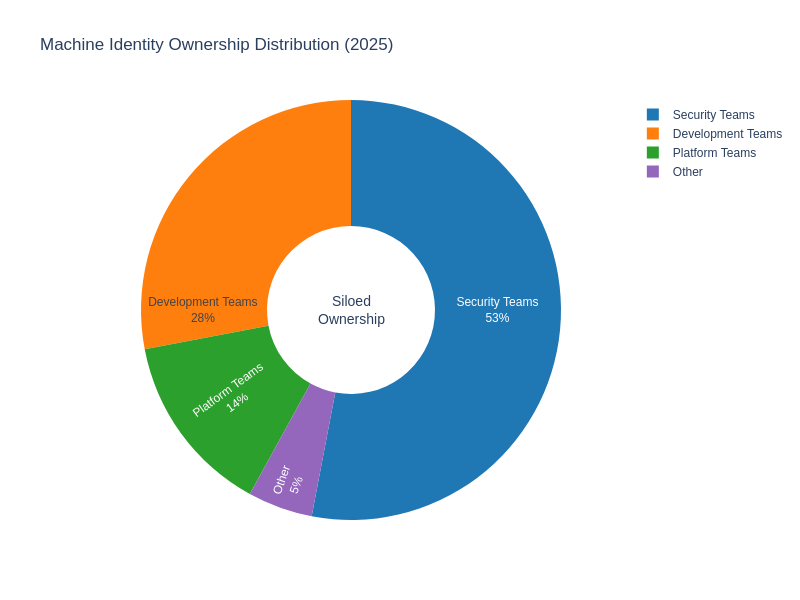}}
\caption{Machine Identity Ownership Distribution (2025). This chart illustrates the fragmented responsibility for machine identity management across organizational teams, with security teams (53\%), development teams (28\%), platform teams (14\%), and others (5\%) sharing ownership, creating potential governance gaps.}
\label{fig6}
\end{figure}

This fragmented ownership creates governance gaps, with 77\% of leaders believing that every undiscovered machine identity represents a potential vulnerability. Other significant challenges include:
\begin{itemize}
\item 74\% are concerned about managing identities in ephemeral cloud workloads
\item 56\% note that IAM teams are responsible for only 44\% of machine identities
\item 37\% report difficulty keeping up with accelerated renewal and rotation requirements
\item 34\% still rely on manual processes for certificate management
\end{itemize}

\begin{figure}[htbp]
\centerline{\includegraphics[width=\columnwidth]{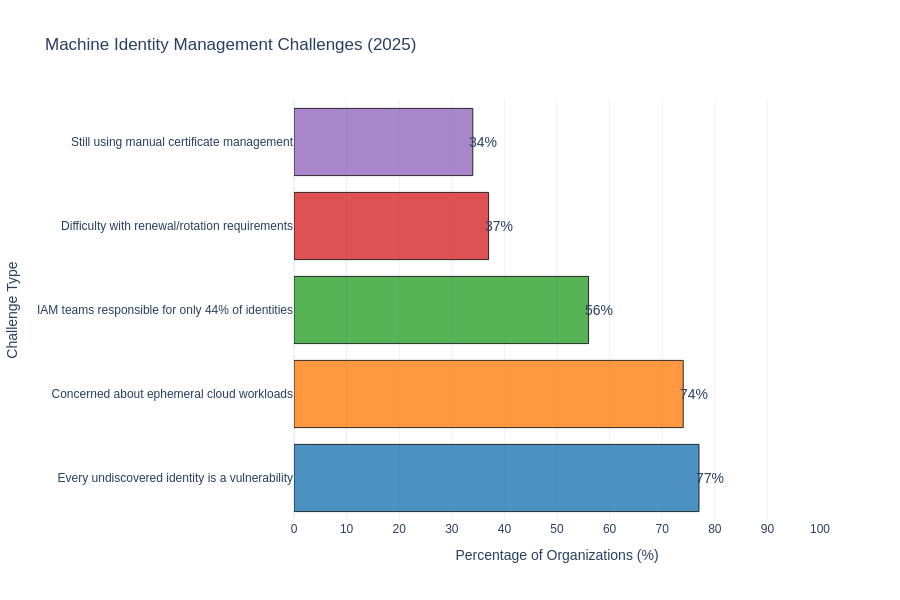}}
\caption{Machine Identity Management Challenges (2025). This figure presents the key operational challenges organizations face in managing machine identities, with ephemeral cloud workload management (74\%) and fragmented responsibility (56\%) emerging as the most significant barriers to effective governance.}
\label{fig7}
\end{figure}

\subsection{Human-Machine Identity Spectrum}
Our analysis led to the development of a human-machine identity spectrum that represents the continuum of identity types in modern enterprises, as shown in Fig. 8.

\begin{figure}[htbp]
\centerline{\includegraphics[width=\columnwidth]{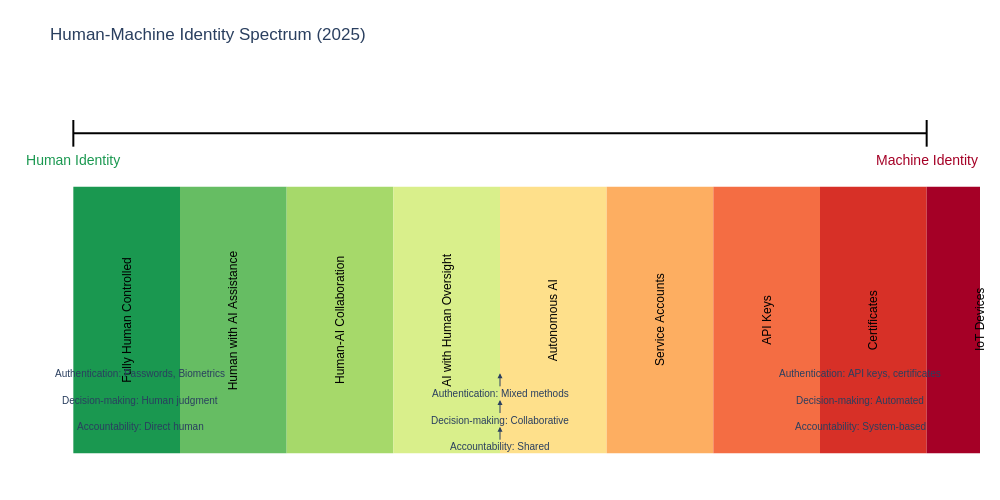}}
\caption{Human-Machine Identity Spectrum (2025). This visualization presents a comprehensive continuum of identity types in modern enterprises, ranging from fully human-controlled identities to autonomous machine identities, with several hybrid forms in between. The spectrum illustrates how traditional binary categorization fails to address the nuanced reality of modern identity management.}
\label{fig8}
\end{figure}

This spectrum ranges from fully human-controlled identities to fully autonomous machine identities, with several hybrid forms in between:
\begin{itemize}
\item Fully Human Controlled: Traditional user accounts with direct human control
\item Human with AI Assistance: Human users leveraging AI tools with limited autonomy
\item Human-AI Collaboration: Shared control between humans and AI systems
\item AI with Human Oversight: AI systems operating with human supervision
\item Autonomous AI: AI systems making independent decisions
\item Service Accounts: Background processes with programmed behavior
\item API Keys: System-to-system communication credentials
\item Certificates: Machine-to-machine trust relationships
\item IoT Devices: Embedded systems with limited human interaction
\end{itemize}

Each point on this spectrum has different authentication methods, decision-making processes, and accountability models. Traditional governance approaches that treat human and machine identities as binary categories fail to address the nuanced reality of this spectrum.

\section{Unified Identity Governance Framework}
Based on our research findings, we propose a Unified Identity Governance Framework that treats all entities accessing systems as part of a continuous spectrum rather than discrete categories.

\subsection{Core Principles}
The unified approach is built on four core principles:

\subsubsection{Identity Continuum vs. Binary Categorization}
Rather than forcing identities into "human" or "machine" buckets, treat identity as a spectrum with varying degrees of human direction and machine autonomy. A fully autonomous AI agent and a human-operated account represent opposite ends of this spectrum, with many hybrid forms in between.

\subsubsection{Risk-Based Approach to All Identities}
Apply consistent risk evaluation across all identity types based on:
\begin{itemize}
\item The sensitivity of resources being accessed
\item The potential impact of compromise
\item The verifiability of the identity
\item The historical behavior patterns
\item The context of access requests
\end{itemize}

\subsubsection{Continuous Verification and Zero Trust Principles}
Apply zero trust principles uniformly across the identity spectrum:
\begin{itemize}
\item Never trust, always verify
\item Verify explicitly based on all available signals
\item Use least privilege access for all identities
\item Assume breach and minimize blast radius
\end{itemize}

\subsubsection{Governance Across the Full Identity Lifecycle}
Implement consistent governance processes for all identities:
\begin{itemize}
\item Creation and provisioning
\item Authentication and authorization
\item Monitoring and analytics
\item Recertification and review
\item Deprovisioning and termination
\end{itemize}

\subsection{Implementation Roadmap}
Implementing unified identity governance requires a structured approach:

\subsubsection{Assessment: Mapping Your Identity Landscape}
Begin by creating a comprehensive inventory of all identities in your environment:
\begin{itemize}
\item Human user accounts
\item Service accounts and application identities
\item API keys and secrets
\item Device and IoT identities
\item AI agent credentials
\item Delegated authority relationships
\end{itemize}

Map the relationships between these identities, particularly where machine identities act on behalf of humans or other machines. Identify governance gaps where certain identity types have less oversight than others.

\subsubsection{Strategy: Developing Unified Policies and Controls}
Create a unified governance framework that applies consistent principles across all identity types:
\begin{itemize}
\item Standardized naming conventions that indicate identity type and purpose
\item Consistent approval workflows for provisioning all identity types
\item Uniform privilege management principles
\item Integrated monitoring and alerting
\item Comprehensive audit and compliance reporting
\end{itemize}

\subsubsection{Technology: Tools and Platforms That Support Unified Governance}
Evaluate and implement technologies that can manage the full spectrum of identities:
\begin{itemize}
\item Identity Governance and Administration (IGA) platforms with machine identity capabilities
\item Privileged Access Management (PAM) solutions that handle both human and machine secrets
\item Cloud Infrastructure Entitlement Management (CIEM) tools
\item API security gateways with strong authentication
\item Behavioral analytics platforms that can baseline both human and machine behaviors
\end{itemize}

\subsubsection{Operations: New Roles and Responsibilities for Identity Teams}
Evolve your organizational structure to eliminate silos between human and machine identity management:
\begin{itemize}
\item Create cross-functional identity governance teams
\item Develop new roles like "Machine Identity Security Architect"
\item Implement joint reviews between IAM, security operations, and development teams
\item Establish clear ownership for hybrid identity scenarios
\end{itemize}

\section{Discussion}
This section discusses the implications of our findings, regulatory considerations, future challenges, and limitations of our proposed framework.

\subsection{Implications for Cybersecurity Practice}
The human-machine identity blur represents a fundamental shift in how organizations must approach identity security. Traditional models that separate human and machine identity management create governance gaps that attackers can exploit.

Our research suggests several key implications for cybersecurity practitioners:
\begin{itemize}
\item Identity teams must evolve to address the full spectrum of identities rather than focusing primarily on human users
\item Security monitoring must incorporate behavioral analytics for both human and machine identities
\item Risk assessment methodologies must account for delegation relationships between humans and machines
\item Incident response playbooks must address hybrid identity attack scenarios
\end{itemize}

Organizations that implement unified governance approaches can expect several benefits:
\begin{itemize}
\item Reduced security incidents through comprehensive visibility
\item Faster incident response through integrated monitoring
\item Improved compliance posture through consistent governance
\item Enhanced operational efficiency through standardized processes
\end{itemize}

\subsection{Regulatory Considerations}
As the field evolves, regulatory frameworks are beginning to address the human-machine identity blur:
\begin{itemize}
\item GDPR and similar privacy regulations increasingly apply to automated decision-making
\item Financial regulations are expanding to cover AI systems acting on behalf of humans
\item Critical infrastructure protection frameworks now include machine identity requirements
\item Industry standards bodies are developing guidance for unified identity governance
\end{itemize}

Organizations should anticipate further regulatory evolution in this area and prepare by implementing comprehensive governance now.

\subsection{Future Challenges}
The identity landscape will continue to evolve rapidly. Organizations should prepare now for emerging trends:

\subsubsection{Preparing for Fully Autonomous AI Agents}
As AI systems become more autonomous, they'll need their own identity governance frameworks:
\begin{itemize}
\item Ethical boundaries and operational constraints
\item Verifiable credentials for AI system capabilities and limitations
\item Audit trails for autonomous decision-making
\item Accountability mechanisms for AI actions
\item Revocation procedures for compromised AI systems
\end{itemize}

Organizations should begin developing governance models for autonomous AI now, rather than waiting for these systems to become widespread.

\subsubsection{Quantum Computing Implications for Identity Security}
Quantum computing threatens many cryptographic foundations of identity systems:
\begin{itemize}
\item Inventory cryptographic algorithms used in identity infrastructure
\item Develop migration plans to quantum-resistant algorithms
\item Implement crypto-agility to quickly respond to new vulnerabilities
\end{itemize}

\subsubsection{Biometric and Neurological Identity Factors}
Advanced biometrics and brain-computer interfaces will further complicate identity:
\begin{itemize}
\item Establish governance for biometric identity data
\item Develop policies for neurological authentication factors
\item Address privacy implications of advanced biometrics
\end{itemize}

\subsection{Limitations of the Proposed Framework}
Our Unified Identity Governance Framework has several limitations:
\begin{itemize}
\item Implementation complexity in large, distributed organizations
\item Potential resistance from teams accustomed to siloed approaches
\item Technology gaps in current identity management platforms
\item Challenges in establishing consistent risk models across diverse identity types
\end{itemize}

Organizations should view the framework as a target state and develop phased implementation plans that address these limitations.

\subsection{Areas for Future Research}
Our work identifies several areas for future research:
\begin{itemize}
\item Formal models for attributing actions in hybrid human-machine systems
\item Standardized risk assessment methodologies for the identity continuum
\item Metrics for measuring the effectiveness of unified governance approaches
\item Ethical frameworks for autonomous AI identity management
\item Regulatory approaches to accountability in human-machine blur scenarios
\end{itemize}

\section{Conclusion}
The human-machine identity blur represents one of the most significant cybersecurity challenges facing organizations today. Our research demonstrates that traditional approaches that treat human and machine identities as separate domains create governance gaps that attackers can exploit.

\subsection{Summary of Key Findings}
Our key findings include:
\begin{itemize}
\item Machine identities now outnumber human identities by a factor of 43:1 in the average enterprise
\item 50\% of organizations experienced security breaches tied to compromised machine identities in the past year
\item Ownership of machine identities is highly fragmented, with IAM teams responsible for less than half
\item Certificate-related outages have increased dramatically, with 45\% of teams reporting weekly incidents
\item The boundary between human and machine identities is increasingly blurred as AI systems act on behalf of humans
\end{itemize}

\subsection{Significance of the Unified Governance Approach}
The Unified Identity Governance Framework addresses these challenges by:
\begin{itemize}
\item Treating identity as a continuum rather than binary categories
\item Applying consistent risk evaluation across all identity types
\item Implementing zero trust principles uniformly
\item Establishing governance across the full identity lifecycle
\end{itemize}

This approach enables organizations to close governance gaps and reduce the risk of identity-based attacks.

\subsection{Practical Recommendations}
Security leaders should take immediate action:
\begin{itemize}
\item Conduct an identity inventory that includes all machine identities
\item Evaluate your IAM program for human-machine boundary blindspots
\item Implement continuous monitoring across all identity types
\item Develop unified governance policies that apply to all entities
\item Train security teams on the unique risks of hybrid identities
\end{itemize}

\subsection{Future Research Directions}
As AI systems become increasingly autonomous, further research is needed on:
\begin{itemize}
\item Governance frameworks for fully autonomous AI agents
\item Attribution models for actions in hybrid human-machine systems
\item Risk assessment methodologies for the identity continuum
\item Regulatory approaches to accountability in human-machine blur scenarios
\end{itemize}

The human-machine identity blur isn't a future concern—it's a present reality that most security programs aren't adequately addressing. Organizations that recognize this blind spot and implement unified identity governance will be better positioned to defend against the sophisticated attacks of 2025 and beyond.

\end{document}